\documentclass{article}

\usepackage[letterpaper,margin=1.0in]{geometry}

\usepackage{todonotes}

\usepackage{amsthm}
\usepackage{physics}
\usepackage{url}

\usepackage{amsmath}
\usepackage[most]{tcolorbox}
\usepackage{slashed}

\usepackage{bm}

\usepackage{comment}

\begin{document}

	\begin{center}
		{\Large \textbf{
				Comment on “Supervised Learning with Quantum Hamiltonian Embeddings” (arXiv:2610.26833) \cite{molteni}
		}}
	\end{center}


\begin{center}
Alexander G.~Abanov$^{1,2}$,
Luca Candelori$^{1}$,
and Kharen Musaelian$^{1}$\\[0.6em]
{\small
$^{1}$Qognitive, Inc., Miami Beach, FL 33139, USA\\
$^{2}$Department of Physics and Astronomy, Stony Brook University,
Stony Brook, NY 11790, USA
}
\end{center}

	\begin{center}
		\today
	\end{center}

In its first version, the paper by Molteni et al.\ \cite{molteni} presented supervised learning with trainable data-dependent Hamiltonians as a new framework, without citing the earlier QCML work. However, the authors could not have been unaware that this idea was introduced by Qognitive as QCML (Quantum Cognition Machine Learning) \cite{musaelian2024,qcmlpatent2025,quantumdevice2025,candelori2026}, based on similar ideas in cognitive science \cite{busemeyer2012}. As a matter of fact, QCML was introduced by Qognitive to the IBM authors of this paper and was the foundation of our shared paper \cite{candelori2026}, in which we successfully ran QCML on IBM's Heron processor for a binary classification task. Moreover, Qognitive built and shared with IBM the codebase for QCML, containing, among other things, a backpropagation implementation identical to that used in this paper. At our request, the authors added a citation to Candelori et al.\ \cite{candelori2026} in version~2 and distinguished their construction from QCML. This acknowledgment does not fully describe the collaborative history outlined here.

The main architectural modification introduced in the current paper is that all training data are encoded in one Hamiltonian and one ground state, while in QCML each data point is encoded with its own Hamiltonian parameterized by that data point and hence corresponds to an input-dependent state, which is essential to the whole method (see, e.g., Ref.~\cite{abanov2025geometry}). The idea of encoding the entire dataset in one Hamiltonian (ground state) was discussed extensively within the Qognitive--IBM collaboration for months, but especially during a working session of Qognitive and IBM on May~17, 2024.

This modification also changes the role of quantum computation at inference. In QCML, a previously unseen input requires a new input-dependent ground-state calculation. In the experimental architecture of Molteni et al., inference is entirely classical. Classical inference does not by itself imply that learning is classically efficient; the question is whether quantum training and the constraints it imposes on possible classical solutions improve predictive accuracy or reduce learning cost relative to suitable classical alternatives. For the experimental architecture, the authors do not compare test accuracy and total training cost with their own classical alternative based on independent local density matrices, which drops global quantum-state compatibility (Sec.~III.B.4).

Below we examine the computational role of the shared state, the trainability and generalization guarantees, and the relevant classical alternatives.

In the hardware implementation, sample-based Krylov quantum diagonalization (SKQD) returns the learned state as an explicit classical list of bitstrings and amplitudes. The quantum processor generates samples to select this support; at fixed support and encoding parameters, the amplitudes are obtained by projected classical diagonalization (Appendix~J). Using standard controlled-gate constructions \cite{barenco1995}, a circuit preparing the same state can be built from these data with a number of gates polynomial in the number of stored bitstrings and qubits. In this implementation, energy and gradient evaluations are also classical. The direct benchmark is therefore classical versus quantum support selection, followed by the same classical state and parameter updates. Although the authors compare SKQD with a classical iterative solver, they do not establish that quantum sampling reduces the total cost of obtaining a useful classifier. The difficulty of unrestricted ground-state preparation does not establish the difficulty of this sparse-state learning problem.

The statement preceding Theorem~1 that the result ``guarantees a useful training direction'' is not justified by the theorem (Sec.~II.B). The theorem gives a lower bound on gradient variance for certain final-layer angles sampled uniformly over a full period, with the state and all other parameters fixed; the gradient at the current angles can still vanish. The proof requires a state with training energy at most $1/2-\gamma$, for $\gamma>0$, and applies equally to classically obtained states satisfying this condition. It bounds neither the cost of obtaining the state nor the number of training iterations. Appendix~C itself acknowledges that absence of barren plateaus does not guarantee successful training.

The generalization theorem (Sec.~III.A) bounds the difference between the training and population Hamiltonians in operator norm, uniformly over the allowed encoding parameters. It therefore controls the energy generalization gap for every normalized state, including one learned directly from the training data. Within the same encoding family, classical training attaining the same training energy receives the same classification guarantee. Neither theorem by itself provides a computational justification for quantum training.

The omission of a benchmark against the authors' own classical alternative in Sec.~III.B.4 is particularly surprising. In their fixed-depth experimental circuit, each single-qubit readout term conjugated by the encoding acts on a fixed, bounded-size region. At fixed encoding parameters, optimizing independent density matrices on these regions requires only small classical eigenproblems. Dropping the requirement that these matrices arise from one global quantum state enlarges the allowed set, so the minimum training cost is no higher than the global ground-state energy. The paper also states that its energy-based generalization argument extends to this relaxation under the stated Lipschitz assumptions. Their separate examples against Pauli surrogates do not address this omission.

In conclusion, the relationship to QCML and the earlier collaborative discussions of the shared Hamiltonian idea deserve explicit acknowledgment. The reported implementation uses quantum-assisted training of a model with entirely classical inference. The paper does not establish any computational advantage over learning this model directly by classical methods. Invoking the hardness of general ground-state problems or a separate cryptographic learning construction does not resolve this deficiency.

\end{document}